\documentclass{article}
\usepackage[utf8]{inputenc}
\usepackage[lmargin=0.65in,rmargin=0.65in,tmargin=0.9in,bmargin=0.75in]{geometry}
\usepackage{hyperref,lmodern,textcomp,graphicx}
\usepackage[table]{xcolor} \usepackage{caption,fancyhdr}
\usepackage{color,amsmath,amssymb,mathpazo,mathtools}
\usepackage{vwcol}
\usepackage{lipsum}
\usepackage{tgpagella}
\usepackage{indentfirst}
\usepackage{bchart}
\usepackage{lettrine,multicol}
\usepackage[sorting=none,style=chem-angew,autocite= superscript]{biblatex}
\usepackage{tcolorbox}%
\definecolor{mycolor}{rgb}{0.122, 0.435, 0.698}
\makeatletter
\newcommand{\mybox}[1]{%
  \setbox0=\hbox{#1}%
  \setlength{\@tempdima}{\dimexpr\wd0+13pt}%
  \begin{tcolorbox}[boxrule=0pt,arc=0pt,
      left=6pt,right=6pt,top=6pt,bottom=6pt,boxsep=1pt,width=\textwidth]
    #1
  \end{tcolorbox}
}

\fancypagestyle{firstpage}
{\fancyhead[L]{\textsc {\color {gray} {\large }}}    
\fancyhead[R]{\large \textsc{\color {gray} {Research Article}}}}
    
\graphicspath{{Figures/}}

\begin{document}

\thispagestyle{firstpage}

{\noindent \LARGE \textit{Designing topological acoustic~lattices via electroacoustic analogies}} \\ [0.5em] 

\noindent {\textit{\large Hasan B. Al Ba'ba'a}$^{1,2,\ddagger}$,} {\large \textit{Kyung Hoon Lee}$^2$,} and {\large \textit{Qiming Wang}$^2$} \\

\noindent \begin{tabular}{c >{\arraybackslash}m{6.5in}}
    \includegraphics[]{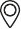} &
    \noindent {\small $^1$Department of Mechanical Engineering, Union College, Schenectady, NY 12308, USA} \\
    
    & \noindent {\small $^2$Sonny Astani Department of Civil and Environmental Engineering, University of Southern~California, Los Angeles, CA 90089, USA} \\[1em]

    \includegraphics[]{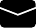}& \noindent {\href{mailto:albabaah@union.edu}{\small albabaah@union.edu}$^\ddagger$}\\
\end{tabular}


\mybox{\textit{Topological acoustics has recently witnessed a spurt in research activity, owing to their unprecedented properties transcending typical wave phenomena.~In recent years, the use of coupled arrays of acoustic chambers has gained popularity in designing topological acoustic systems.~In their common form, an array of acoustic chambers with relatively large volume is coupled via narrow channels. Such configuration is generally modeled as a full three-dimensional system, requiring extended computational time for simulating its harmonic response. To this end, this paper establishes a comprehensive mathematical treatment of the use of electroacoustic analogies for designing topological acoustic lattices. The potential of such analytical approach is demonstrated via two types of topological systems:~(i) edge states with quantized winding numbers in an acoustic diatomic lattice and (ii) valley Hall transition in an acoustic honeycomb lattice that leads to robust waveguiding. In both cases, the established analytical approach exhibits an excellent agreement with the full three-dimensional model, whether in dispersion analyses or the response of an acoustic system with a finite number of cells.~The established analytical framework is invaluable for designing a variety of acoustic topological insulators with minimal computational cost. }
}

\noindent \textbf{Keywords
}

\noindent Topological Insulators; Acoustic Lattices; Electroacoustic Analogies; Quantum Valley Hall Effect; Edge states.

\noindent\rule{7.2in}{0.5pt}

\section{Introduction}
The study of topological acoustics has recently witnessed a spurt in research activity, owing to their unprecedented properties ranging from robustness against defects and wave unidirectionality to backscattering-immunity \cite{Zhang2018TopologicalSound,Ma2019TopologicalSystems}. Such interesting behaviors enabled various applications with inherent topological protection, such as robust wave guiding \cite{Lu2017ObservationCrystals}, one-way signal transport \cite{He2016AcousticTransport}, logic operations \cite{Xia2018ProgrammableInsulator}, and negative refraction \cite{He2018TopologicalCrystal}.~Innovative topological acoustic systems have been advancing in parallel to topological elastic structures, offering elastic realizations of a wide range of topological properties \cite{Vila2017ObservationLattice,liu2018tunable,zhu2018design,chaunsali2018subwavelength,Serra-Garcia2018ObservationInsulator,chen2019mechanical,fan2019elastic}.

In recent years, the use of coupled arrays of acoustic chambers has gained popularity in designing acoustic topological insulators \cite{ni2017topological,Khanikaev2015TopologicallyLattice,Yang2018AcousticSystem,gao2020acoustic,ni2019observation,Xue2019AcousticLattice,Qi2020AcousticInsulators}. In their common form, acoustic chambers with a relatively large volume are coupled with their neighboring chambers via narrow channels in one or more directions, henceforth referred to as \textit{acoustic lattices}. The fascinating part of using acoustic lattices is their ability to closely mimic Hamiltonians pertaining to a variety of topological electronic systems. To date, acoustic lattices have been utilized in inducing robust edge states \cite{ni2017topological,Khanikaev2015TopologicallyLattice}, Quantum Valley Hall Effect (QVHE) \cite{Yang2018AcousticSystem,gao2020acoustic}, Quantum Spin Hall Effect (QSHE) \cite{gao2020acoustic}, Hofstadter-butterfly effect in quasi-periodic lattices \cite{ni2019observation} and, more recently, higher order topological insulators with corner states \cite{Xue2019AcousticLattice,Qi2020AcousticInsulators,Ni2020DemonstrationInsulator}. 

Electroacoustic analogies (i.e., mapping acoustic systems to circuitry) have been demonstrated as a suitable methodology for studying acoustic systems \cite{nouh2014onset,Zhang2010AcousticApplications,roshwalb2014performance,akl2010multi,akl2012multicell}. Established circuit analogues typically consist of a series of coupled inductors and capacitors, whose inductance and capacitance are estimated based on the geometrical and medium properties of an acoustic system. While effective Hamiltonians are commonly established via parameter fitting of the coupling in an acoustic lattice, owing to their three-dimensional nature \cite{ni2017topological}, the use of electroacoustic analogies in modeling Hamiltonian of an acoustic Kagome lattice has been recently demonstrated to ultimately design topological corner modes \cite{guan2021method}.

In light of the aforementioned studies, this work aims to establish a comprehensive analytical framework for acoustic lattices using electroacoustic analogies to design non-trivial topological states. The key importance of such approach is two-fold: (i) to provide an analytical insight into the topological protection mechanism in acoustic lattices and (ii) to reduce computational cost for analyzing them. While modeling acoustic lattices via harnessing lumped circuit equivalents is convenient in terms of fast calculation and obtaining analytical models, the approach becomes less accurate at relatively higher frequencies \cite{Zhang2010AcousticApplications,guan2021method}. Nonetheless, as will be shown, the first few dispersion branches and their corresponding frequency response in a finite acoustic lattice can be estimated with relatively high accuracy, rendering such a methodology an asset in studying topological acoustics with high computational efficiency.

To illustrate the concept, an acoustic monatomic lattice {is studied as a starting point} to establish variables and physical meaning of coupled chambers. Parameters of the dispersion relations are interpreted based on the renowned Helmholtz resonance, which enables a better estimation of resonance frequencies based on the geometry of the acoustic system. Afterwards, two types of topological acoustic lattices are investigated: (i) a diatomic lattice with non-vanishing quantized winding number and edge states, and (ii) a honeycomb lattice with QVHE for designing robust waveguides. The developed circuit model for both lattices is compared with the numerical results from a full-scale three-dimensional counterpart for unit cell and finite lattice analyses. 

\section{Acoustic monatomic lattices}

\subsection{Electroacoustic analogies: Modeling and geometrical dependence}
As a starting point, the electroacoustic analogy is demonstrated in a chain of identical acoustic chambers (or cavities) coupled via narrow channels with rigid walls to satisfy sound-hard boundaries (Fig.~1(\textit{a,b})). We shall refer to this chain as an \textit{acoustic monatomic lattice} due to its resemblance to a typical monatomic lattice as will be proven shortly. At relatively low frequencies, the acoustic pressure inside each of these chambers is assumed to be constant throughout, thus constituting its only degree of freedom, denoted here as $p_i$ with $i$ symbolizing the chamber's order in the acoustic chain (Fig.~1(\textit{a})). As a result, electroacoustic analogies can be readily utilized and offer an intriguing mapping of the acoustic pressure and particle velocity in the acoustic medium to voltage and current in circuitry \cite{Zhang2010AcousticApplications}. In fact, narrow channels and acoustic chambers are analogous to electric inductors with inductance $L_0$ and grounded capacitors of capacitance $C_0$, respectively, and their values are estimated based on chamber/channel geometrical properties \cite{Zhang2010AcousticApplications}:
\begin{subequations}
\begin{align}
L_0& =\frac{\rho\ell_e}{S_c}\\
    C_0& =\frac{V}{\rho c^2}
\end{align}
\label{eq:ind_cap_def}
\end{subequations}
    Here, $S_c$ is the cross-sectional area of the narrow channel, $\rho$ is the fluid's density, $V$ is the chamber’s volume, $c$ is the sound speed in the fluid medium, and $\ell_e$ is the effective length of air channels (which is often longer than the channel’s physical length $\ell$). Throughout this paper, air is chosen as the fluid medium and its properties are listed in Table~\ref{table:air_properties}.

\begin{table*}[h]
\caption{Air properties used in simulations.}
\centering
\begin{tabular}{l l l l c c}
\hline
Property & Temperature & Density ($\rho$) & Sonic Speed ($c$) \\
\hline
Value & 25 $^\text{o} \text{C}$ & 1.1842 kg/$\text{m}^3$ & 346.12 m/s \\
\hline
\end{tabular}
\label{table:air_properties}
\end{table*}

\subsection{Dispersion relation and relevance to Helmholtz resonance}

Establishing an equivalent circuit model to the acoustic monatomic lattice is intended to derive a simplified analytical expression of its dispersion relation to avoid analyzing a full-scale three-dimensional unit cell. Assuming harmonic motion, electrical current $I$ flowing through an inductor (a capacitor), in response to a drop in voltage $\Delta V$, is governed by the reciprocal of a frequency-dependent impedance $Z_0=\textbf{i} \omega L_0$ ($Z_0=1/(\mathbf{i} \omega C_0)$), yielding $I=\Delta V/Z_0$ ($\mathbf{i}$ is the imaginary unit).~Exploiting the mapping of acoustic pressure to voltage and knowing that currents entering and exiting a junction sum to zero (per Kirchhoff law), the governing equation of the $i^{\text{th}}$ acoustic chamber is:
\begin{equation}
    \left(\frac{2}{\mathbf{i}\omega L_0} +\mathbf{i}\omega C_0 \right) p_{i}- \frac{1}{\mathbf{i}\omega L_0} (p_{i+1}+p_{i-1})=0
\end{equation}
Applying Bloch theorem $p_{i\pm1}=p_i e^{\pm\mathbf{i}\mu}$ and multiplying by $\mathbf{i}\omega L_0$, further symbolic simplifications yield the dispersion relation:
\begin{equation}
    \omega = 2\omega_0 \left|\sin \left(\frac{\mu}{2} \right) \right|
    \label{eq:ML_disp}
\end{equation}
where $\mu$ is the non-dimensional wavenumber and:
\begin{equation}
    \omega_0 = \sqrt{\frac{1}{L_0 C_0}}
    \label{eq:Hel_resonance}
\end{equation}
Equation (\ref{eq:ML_disp}) is in perfect analogy with the dispersion relation of a typical elastic monatomic lattice, albeit with the definition of $\omega_0$ being the natural frequency of a spring-mass system \cite{albabaa2018TDF}. For our acoustic lattice, the quantity $\omega_0$ in Equation (\ref{eq:Hel_resonance}) is interpreted as the natural frequency of a stand-alone Helmholtz resonator (an isolated unit cell as seen in Fig. 1(\textit{b})), owing to the fact that a Helmholtz resonator is often mapped to a spring-mass system \cite{Yuan2007ActiveImpedance,Alster1972ImprovedResonators,Xu2010DualResonator}. Consequently, an acoustic monatomic lattice can be perceived as serially coupled Helmholtz resonators. Implementing this physical understanding, in addition, facilitates a better estimation of acoustic lattice properties. For example, the effective length $\ell_e$ for the neck of a Helmholtz resonator (i.e., air channel) is suggested to be:
\begin{equation}
    \ell_e = \ell(1+\delta_c)
    \label{eq:eff_length}
\end{equation}
where the correction factor $\delta_c$ is the summation of the inner and outer end corrections of the channel's length, which are chosen here as $\delta_{\text{in}}=0.425d_h/\ell$ and $\delta_{\text{out}}=0.3d_h/\ell$, respectively, for flanged and unflanged ends \cite{Raichel2006TheAcoustics}. Here, $d_h=4S_c/P_c$ is the channel’s hydraulic diameter and $P_c$ denotes the channel’s perimeter \cite{wibulswas1966laminar}. Accordingly, the complete correction factor is $\delta_c= \delta_{\text{in}} + \delta_{\text{out}}= 0.725d_h/\ell$. Note that the hydraulic diameter for a square (circular) channel is equal to the channel's side length (nominal diameter). Making use of Equations (\ref{eq:ind_cap_def}), (\ref{eq:Hel_resonance}), and (\ref{eq:eff_length}), the Helmholtz resonance can be explicitly expressed in terms of the properties of the acoustic medium and resonator's geometry (in units of Hertz):
\begin{equation}
    f_\text{R} = \frac{\omega_0}{2\pi} =  \frac{c}{2\pi\ell} \sqrt{\frac{V_c/V}{1+\delta_c}}
    \label{eq:hel_resonance_Hz}
\end{equation}
where $V_c$ is the volume of the channel. Equation (\ref{eq:hel_resonance_Hz}) implies that the larger the ratio of the volumes, the larger the natural frequency, given a constant length $\ell$ (Fig.~1(\textit{c})). Doubling this quantity provides an estimate for the cutoff frequency (in Hertz) of the acoustic monatomic lattice, as inferred from Equation (\ref{eq:ML_disp}).

\begin{figure*}[]
     \centering
\includegraphics[]{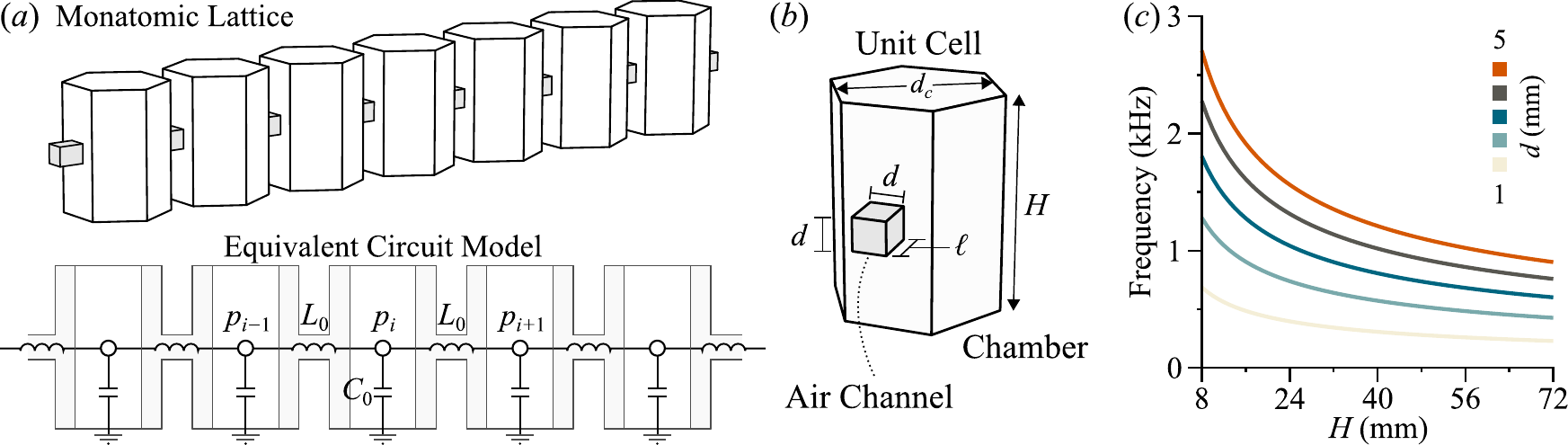}
     \caption{(\textit{a}) Schematic of an acoustic monatomic lattice and its equivalent electrical circuit model. (\textit{b}) Unit cell of the acoustic monatomic lattice detailing the geometrical properties of the chamber and channel. (\textit{c}) Fundamental frequency of an isolated unit cell (acting as a Helmholtz resonator) with varying chamber height $H$ and square channel side $d$. The length of the air channel and the diameter of the hexagon’s inscribed circle remain unchanged and are chosen to be $\ell=4$mm and $d_c=14$mm, respectively.}
     \label{fig:monatomiclattice}
\end{figure*}

\subsection{Numerical examples}
Three different combinations of the acoustic chamber/channel’s geometrical cross-sectional areas are simulated: (1) square chambers with square channels, (2) hexagonal chambers with square channels, and, (3) circular chambers with circular channels (Fig.~2). These choices of geometrical shapes are intended to demonstrate the ability of the reduced analytical model to predict the dispersion relation with relatively high accuracy regardless of the general geometrical shape of both chamber and channel. In all cases, three-dimensional model simulations are achieved via COMSOL commercial software and assuming a discretization of the Quadratic Lagrange type. For distinction between the circuit model and the full three-dimensional model, the dispersion relation or related system dynamics obtained from the circuit model (COMSOL) is labeled as \textit{analytical} (\textit{numerical}). Calculating the dispersion relation for a swept range of the hydraulic diameter $d$ and height $H$, while maintaining the length $\ell=4$mm and parameter $d_c=14$mm (from which the area $S_c$ is calculated) constant throughout, it is seen that all analytical dispersion relations closely resemble their numerical counterparts (Fig.~2). As such, it is now evident that Equation~(\ref{eq:hel_resonance_Hz}) can effectively estimate the natural frequency of the stand-alone Helmholtz resonator with excellent accuracy, and consequently producing the most fitting dispersion relation per Equation~(\ref{eq:ML_disp}). This also displays the competence of the circuit model in predicting the dispersion relation while maintaining very low computational cost relative to the three-dimensional model. Finally, it is important to re-iterate that the geometry of chamber (i.e., square, hexagon or circle) and channels (i.e., square or circle) has no direct influence on the functionality of the acoustic system at relatively low frequencies, and can be fully characterized as a lumped-parameter circuit model, thus further emphasizing the simplicity of the approach adopted here.

\begin{figure*}[]
     \centering
\includegraphics[]{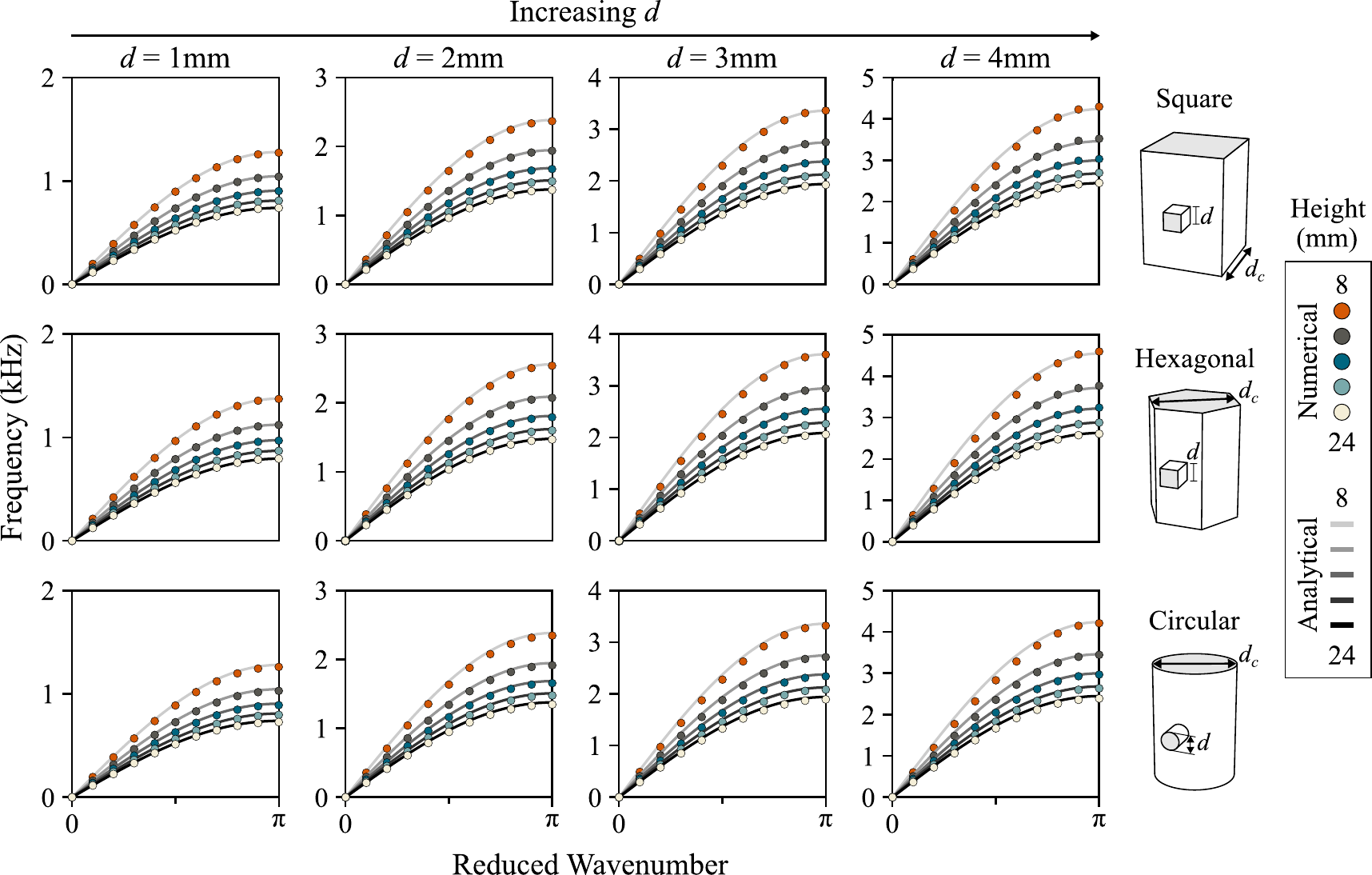}
     \caption{Dispersion relations of acoustic monatomic lattices with different combinations of parameters and geometrical shapes. As shown in the rightmost panel, parameter~$d_c$ ($d$) defines the necessary dimension to calculate the cross-sectional area of the different chamber (channel) shapes, which extends a distance of $H$ ($\ell$).~The analytically obtained dispersion relation from the electrical acoustic model (color-coded lines) are plotted against the full-scale three-dimensional model simulated via commercial software COMSOL (color-coded circles) for different heights $H$, showing excellent agreement. In all cases, the channel length $\ell$ and parameter $d_c$ remain constant throughout and are chosen to be 4mm and 12mm, respectively.}
     \label{fig:monatomiclattice_disp}
\end{figure*}

\section{Acoustic diatomic lattices}
\subsection{Mathematical formulation and dispersion analysis}

Next, an acoustic diatomic lattice constituting an array of equally sized acoustic chambers, coupled via alternating narrow/wide channels with hydraulic diameters of $d_1$ and $d_2$ is considered (Fig.~3).~The emergence of edge states is demonstrated via analyzing the band topology associated with the unit cell of the diatomic lattice.~Analogous to the analysis of the acoustic monatomic lattice, the governing equations of a diatomic unit cell are derived:
\begin{subequations}
\begin{equation}
    \left(\frac{1}{\mathbf{i}\omega L_1} +\frac{1}{\mathbf{i}\omega L_2}+\mathbf{i}\omega C \right) p_i-\left(\frac{1}{\mathbf{i}\omega L_1} q_i+\frac{1}{\mathbf{i}\omega L_2} q_{i-1} \right)=0
\end{equation}
\begin{equation}
\left(\frac{1}{\mathbf{i}\omega L_1} +\frac{1}{\mathbf{i}\omega L_2}+\mathbf{i}\omega C \right) q_i-\left(\frac{1}{\mathbf{i}\omega L_1} p_i+\frac{1}{\mathbf{i}\omega L_2} p_{i+1} \right)=0
\end{equation}
\label{eq:DL_EOM}
\end{subequations}
where $C$ is the capacitance of chambers and $L_{1}$ ($L_{2}$) is the inductance of the first (second) Helmholtz resonator of the diatomic unit cell.~The acoustic pressure in the first (second) chamber of the $i^{\text{th}}$ unit cell is denoted by $p_i$ ($q_i$). Introducing $\omega_1=1/\sqrt{CL_{1}}$ ($\omega_2=1/\sqrt{CL_{2}}$) as the fundamental frequency of the first (second) Helmholtz resonator, the governing equations in (\ref{eq:DL_EOM}) are further simplified as follows:
\begin{subequations}
\begin{equation}
    \left(\omega^2_1 +  \omega^2_2-\omega^2\right) p_i-\left(\omega^2_1 q_i+\omega^2_2 q_{i-1} \right)=0
\end{equation}
\begin{equation}
 \left(\omega^2_1 +  \omega^2_2-\omega^2\right) q_i-\left(\omega^2_1 p_i+\omega^2_2 p_{i+1} \right)=0
\end{equation}
\end{subequations}
Finally, applying the Bloch theorem yields an eigenvalue problem:
\begin{equation}
    \mathbf{H}\mathbf{p}_i = \omega^2 \mathbf{p}_i 
    \label{eq:EVP_DL}
\end{equation}
where the unit-cell’s pressure vector $\mathbf{p}_i$ and Hamiltonian $\mathbf{H}$, respectively, are: 
\begin{subequations}
\begin{equation}
    \mathbf{p}_i^\text{T}= \{ p_i \ q_i\}
\end{equation}
\begin{equation}
    \mathbf{H} = 
    \begin{bmatrix}
    \omega_1^2+\omega_2^2 & -\omega_1^2-\omega_2^2 \text{e}^{-\mathbf{i}\mu} \\
    -\omega_1^2-\omega_2^2 \text{e}^{\mathbf{i}\mu} & \omega_1^2+\omega_2^2
    \end{bmatrix}
\end{equation}
\end{subequations}
It is straightforward to compute the eigenvalues from $|\mathbf{H}-\omega^2 \mathbf{I}|=0$, which results in the dispersion branches:
\begin{equation}
    \omega = \sqrt{\omega^2_1 + \omega^2_2 \pm \sqrt{\omega^4_1 + \omega^4_2 + 2 \omega^2_1 \omega^2_2 \cos(\mu)}}
    \label{eq:DL_disp}
\end{equation}
\begin{figure*}[]
     \centering
\includegraphics[]{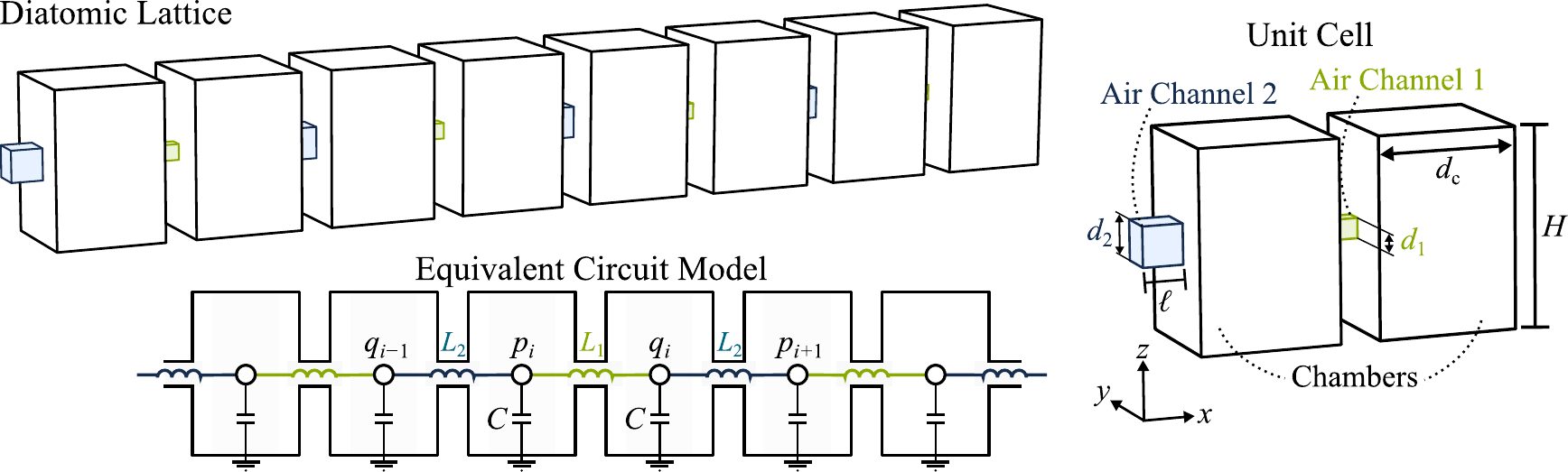}
     \caption{Schematics of an acoustic diatomic lattice, its equivalent electrical circuit model, and its unit cell definition with detailed geometrical properties. Note that the chambers (channels) are of square cross-sectional area with sides $d_c$ ($d_{1,2}$) and the chamber dimensions are identical throughout the acoustic chain~(i.e., constant height $H$ and side length $d_c$).}
     \label{fig:DiatomicLattice}
\end{figure*}

Figure 4(\textit{a}) shows the analytical dispersion branches that are computed directly from Equation (\ref{eq:DL_disp}) by sweeping the values of the wavenumber in the Brillouin zone, i.e., $\mu \in [-\pi,\pi]$, for the cases of (i) $\omega_2<\omega_1$ , (ii) $\omega_2=\omega_1$ and (iii) $\omega_2>\omega_1$. Superimposed are the numerically obtained dispersion relation (presented as circles), displaying excellent agreement with their analytical counterpart (presented as lines).~Herein, the nominal length of the air channel $\ell$ is kept at $4$mm and the values of $\omega_{1,2}$ are changed by introducing the parameterization $d_{1,2}=d_0\mp \Delta d/2$, where $\Delta d$ is the difference in the channels' hydraulic diameter and $d_0=(d_1+d_2)/2$ is their average and chosen here as $d_0 = 3$mm. In this example, the difference in the hydraulic diameter is $\Delta d=-1,0,+1$ mm, which gives (i) $d_1=3.5$mm and $d_2=2.5$mm, (ii) $d_1=d_2=3$mm, and (iii) $d_1=2.5$mm and $d_2=3.5$mm, respectively. Chamber’s height and side length (diameter) are $H=20$mm and $d_c=14$mm, respectively, and they remain unchanged throughout. As expected, a bandgap opens only for $\omega_2\neq \omega_1$ and its limits are $\omega = \sqrt{2} \omega_1$ and $\omega = \sqrt{2} \omega_2$, which are computed from the dispersion relation at the Brillouin zone edges $\mu = \pm \pi$ (Fig. 4(\textit{b})). Moreover, the first (Fig. 4(\textit{a})\textbf{i}) and last (Fig. 4(\textit{a})\textbf{iii}) dispersion relations have no apparent difference as they signify a different choice of the unit cell by flipping the order of narrow/wide channels, which warrants an identical dispersion relation as evident from Equation (\ref{eq:DL_disp}). Note that the order of the bandgap limits is contingent on $\omega_{1,2}$ values and, hence, its bandgap width is $\Delta \omega = \sqrt{2} |\omega_1-\omega_2|$.

\begin{figure*}[ht]
     \centering
\includegraphics[width=0.98\textwidth]{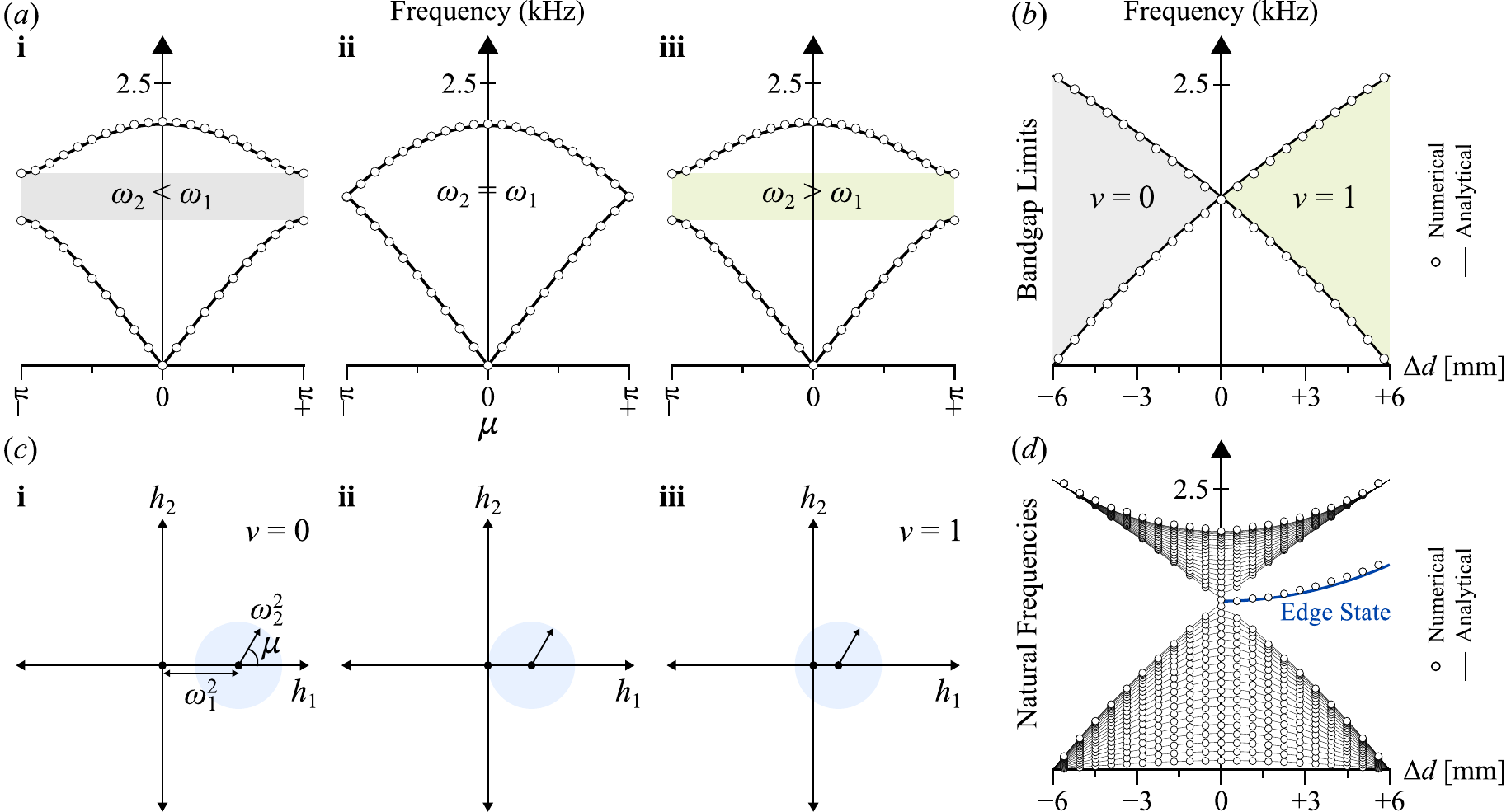}
     \caption{(\textit{a}) Dispersion relations for acoustic diatomic lattices with the following three combinations of the Helmholtz resonance frequencies: (\textbf{i}) $\omega_2<\omega_1$, (\textbf{ii}) $\omega_2=\omega_1$, and (\textbf{iii}) $\omega_2>\omega_1$, corresponding to $\Delta d$ being $-1$mm, 0, and $+1$mm, respectively. (\textit{b})~Bandgap limits for a swept range of $\Delta d \in[-6\text{mm},+6\text{mm}]$, showing the transition of the winding number from being $\nu = 0$ (i.e., topologically trivial) to $\nu = 1$ (i.e., topologically non-trivial) with the bandgap closing at $\omega_2=\omega_1$ (i.e., $\Delta d = 0$). Note that the winding number does not change as long as the bandgap remains open.~(\textit{c}) Functions $h_1$ versus $h_2$ from Equations~(\ref{eq:H_parameters_DL1})~and~(\ref{eq:H_parameters_DL2}), respectively, showing the corresponding winding number associated with their oriented path for (\textbf{i}) $\omega_2<\omega_1$, (\textbf{ii}) $\omega_2=\omega_1$, and (\textbf{iii}), similar to elastic diatomic lattices \cite{chen2018topological}.~(\textit{d}) Natural-frequency distribution of a finite acoustic chain of 40 chambers with open-open boundary conditions, showing the emergence of doubly degenerate edge states when the winding number is non-zero. All simulations show an excellent agreement between analytical and numerical results.}
     \label{fig:DiatomicLattice_disp}
\end{figure*}

\begin{figure*}[ht]
     \centering
\includegraphics[]{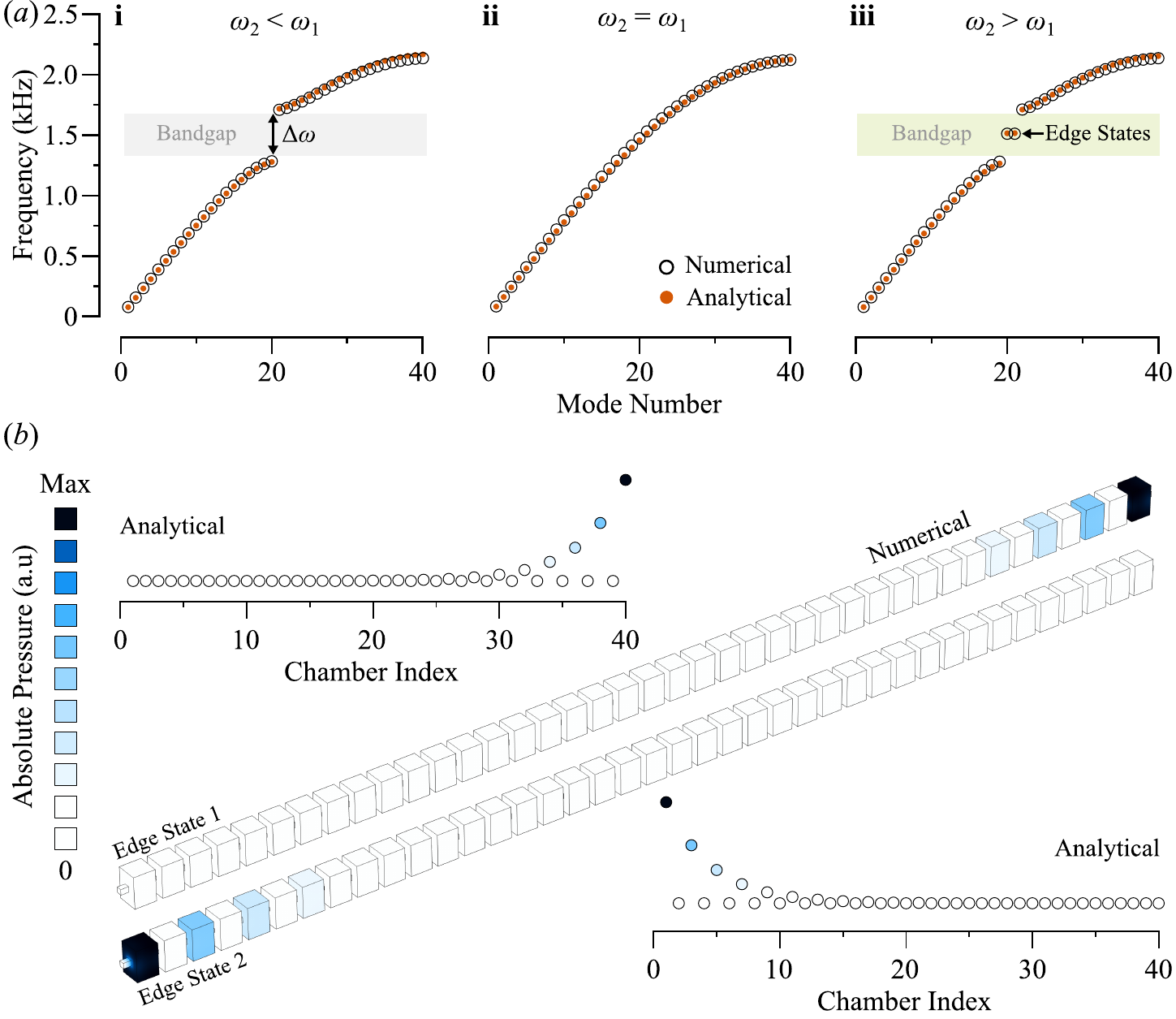}
     \caption{(\textit{a})~Natural-frequency spectrum for three different acoustic diatomic lattices with (\textbf{i}) $\omega_2 < \omega_1$, (\textbf{ii}) $\omega_2 = \omega_1$, and (\textbf{iii}) $\omega_2 > \omega_1$, showing the emergence of edge states only when $\omega_2>\omega_1$ in agreement with the non-zero winding number in Fig.~4(\textit{b,c}).~(\textit{b})~Comparison between the mode shapes of the topological edge states obtained from the full-scale three-dimensional and analytical circuit models.}
     \label{fig:DiatomicLattice_finite}
\end{figure*}

\subsection{Winding number and edge states}
\label{sec:DL_edge_states}
Although the dispersion relations in Figs.~4(\textit{a})\textbf{i},\textbf{iii} are identical, a deeper look into their band topology unravels their distinct topological nature.~To attain an insight into the topological properties of the lattice, the Hamiltonian is recast as a summation of Pauli matrices:
\begin{equation}
    \mathbf{H} = -\sum_{l}h_l \boldsymbol{\sigma}_l
    \label{eq:Pauli_matrices}
\end{equation}
where $l=0,1,2,3$ and
\begin{subequations}
\begin{align}
    h_0 &= - (\omega_1^2 + \omega_2^2) \\
    \label{eq:H_parameters_DL1}
    h_1 &= \omega_1^2 + \omega_2^2 \cos(\mu) \\
    \label{eq:H_parameters_DL2}
    h_2 &= \omega_2^2 \sin(\mu) \\
    h_3 &= 0
\end{align}
\label{eq:H_parameters_DL}
\end{subequations}
Note here that the zeroth Pauli matrix $\boldsymbol{\sigma}_0$ is a $2\times2$ identity matrix. In this class of periodic systems, a quantized winding number, denoted as $\nu$, is sought to achieve topologically-protected edge states. To guarantee such quantization, the diagonal of the Hamiltonian for the diatomic chain must be constant, suggesting that $\boldsymbol{\sigma}_3$ does not play any role in the Hamiltonian \cite{Asboth2016AInsulators}. This condition is already satisfied by setting equal capacitance $C$, which ultimately returns $h_3=0$ as evident from Equation~(\ref{eq:H_parameters_DL}). The meaning of the winding number can be interpreted graphically from plotting $h_1$ versus $h_2$, as functions of $\mu$, which gives a perfect circle of radius $\omega_2^2$ with its center shifted by $\omega_1^2$ from the origin. If the depicted circle winds the origin once, the winding number $\nu$ shall be unity (as in the case of $\omega_2>\omega_1$), and it vanishes otherwise (as in the case of $\omega_2<\omega_1$). A topological transition occurs when $\omega_2=\omega_1$ such that the circle touches the origin and the winding number becomes undefined (Fig.~4(\textit{c})). It is rather interesting to note that the elastic diatomic lattice has an identical behavior to that observed in Fig.~4(\textit{c}), albeit that the circle radius and distance from the origin are related to the lattice's spring constants \cite{chen2018topological}. 

A quantized winding number dictates the emergence, or lack thereof, of topological edge states at the boundaries of a chain with a finite number of unit cells, $n$.~As the boundary of a finite chain is well-defined, unlike the infinite case where the structure is theoretically unbounded, the distinction between the cases of $\omega_2>\omega_1$ or $\omega_2<\omega_1$ is unambiguous. Therefore, a finite acoustic diatomic lattice that is terminated at channels with hydraulic diameter of $d_2$ from both ends is considered. The boundary conditions at both ends are open, i.e., the pressure of peripheral chambers is connected to a node with zero oscillatory pressure via an $L_2$ inductor. Following a similar parametrization procedure to that in the unit cell analysis, the unforced response of the lattice, in frequency domain, is governed by:
\begin{equation}
\begin{bmatrix}
\mathbf{D}-\omega^2\mathbf{I}
\end{bmatrix}\mathbf{p}(\omega) = \mathbf{0}
\end{equation}
where the degrees of freedom are the acoustic pressure in chambers and read:
\begin{equation}
    \mathbf{p}(\omega) = 
    \begin{Bmatrix}
    p_1 & q_1 & p_2 & q_2 & \dots & p_n & q_n
    \end{Bmatrix}^\text{T}
\end{equation}
The matrix $\mathbf{D}$ is a tridiagonal 2-Toeplitz matrix of size $2n\times2n$ that dictates the dynamical characteristics of the acoustic system and is solely a function of $\omega_{1,2}$:
\begin{equation}
    \mathbf{D} = (\omega^2_1 + \omega^2_2)\mathbf{I}_{2n}-\mathbf{\Omega}
\end{equation}
where $\mathbf{I}_{[\cdot]}$ is an identity matrix with its size indicated in the subscript and
\begin{equation}
\mathbf{\Omega} = 
    \begin{bmatrix}
    0 & \omega_1^2 & 0 & \cdots & 0 \\
    \omega_1^2 & 0 & \omega_2^2 & \ddots & \vdots \\
    0 &\omega_2^2 & \ddots & \ddots & 0 \\
    \vdots & \ddots & \ddots & \ddots & \omega_1^2 \\
    0 & \dots & 0 & \omega_1^2 & 0 \\
    \end{bmatrix}
\end{equation}
It is worth noting that the structure of $\mathbf{D}$ resembles the dynamical matrix of a fixed-fixed elastic diatomic lattice with constant masses and distinct stiffness coefficients \cite{al2017pole}. Numerically solving the eigenvalues of matrix $\mathbf{D}$ for swept values of $\Delta d$ (thus simultaneously changing $\omega_{1}$ and $\omega_{2}$) reveals the emergence of repeated edge states, which are pinned at:
\begin{equation}
    \omega = \sqrt{\omega_{1}^2+\omega_{2}^2}
    \label{eq:DL_edgestate}
\end{equation}
only when $d_2>d_1$ (or $\omega_2>\omega_1$). The latter is expected in accordance to the non-vanishing winding number from band topology predictions, as depicted in Fig.~4(\textit{b}). 

An analytical proof of the emergence of edge states at the frequency in Equation~(\ref{eq:DL_edgestate}) can be obtained by examining the characteristic equation of matrix $\mathbf{D}$. Following the developments in \cite{al2017pole,da2007characteristic}, $\mathbf{D}$ is categorized as an unperturbed 2-Toeplitz matrix and its characteristic equation can be shown to be:
\begin{equation}
\frac{1}{\sin(\mu)} \left(
\sin((n+1)\mu)+\frac{\omega^2_2}{\omega^2_1}\sin(n\mu) \right)= 0
\label{eq:mu_analytical}
\end{equation}
The solutions of $\mu$ that satisfy Equation~(\ref{eq:mu_analytical}) are key for finding the natural frequencies, which are computed by substituting the solutions of $\mu$ back into the dispersion relation in Equation~(\ref{eq:DL_disp}). Of interest here is to derive an analytical expression for the edge states within the bandgap. The wavenumber in such a case is a complex number $\mu = \mu_\text{R} + \mathbf{i} \mu_\text{I}$, and the real component of the wavenumber has a magnitude of $\pi$. For a sufficiently large lattice, setting $n \rightarrow \infty$ is a feasible assumption and Equation~(\ref{eq:mu_analytical}) reduces to:
\begin{equation}
\text{e}^{\mu_\text{I}} - \frac{\omega^2_2}{\omega^2_1} = 0
\label{eq:reduced_ChEq_n_infty}
\end{equation}
An analytical expression for the imaginary component of the wavenumber, i.e., $\mu_\text{I}$, can be found via the dispersion relation in (\ref{eq:DL_disp}) and is given by:
\begin{equation}
    \mu_\text{I} = \Im \left( \cos^{-1} \left( \Phi\right) \right) = - \ln \left(\Phi + \sqrt{\Phi^2 - 1} \right)
    \label{eq:muI}
\end{equation}
where 
\begin{equation}
    \Phi = \frac{\omega^4}{2\omega^2_1 \omega^2_2} - \frac{(\omega^2_1+\omega^2_2)\omega^2}{\omega^2_1 \omega^2_2} +1 
    \label{eq:Phi}
\end{equation}
Substituting Equations~(\ref{eq:muI})~and~(\ref{eq:Phi}) back into (\ref{eq:reduced_ChEq_n_infty}), the resulting characteristic equation boils down to:
\begin{equation}
    \left(\omega^2 - (\omega^2_1+\omega^2_2)^2 \right)^2 = 0
\end{equation}
which serves as a proof for the existence of double roots at the frequency in Equation~(\ref{eq:DL_edgestate}).~Interestingly, such a frequency exhibits the largest attenuation coefficient within the bandgap as it is a root for $\partial \Phi/\partial \omega = 0$ \cite{al2017pole}. 

Next, the analytical predictions are compared to the numerical ones (obtained from COMSOL simulations), as depicted in Fig.~4(\textit{d}). To avoid discrepancies in computed natural frequencies in the numerical problem, the length of both open-end channels must be adjusted by considering the effective length $\ell_e$ rather than the nominal value $\ell$. Recall that the effective length for a Helmholtz resonator’s neck is computed by adding two corrections: one for the inner side (connected to the chamber with correction factor $\delta_\text{in}=0.425d_h/\ell$) and one for the outer side (open end with correction factor $\delta_\text{out}=0.3d_h/\ell$).~As such, the correction for end channels needs to be modified only for the outer (open) end, given that the inner end correction is already accounted for in the three-dimensional model.~That is, the open channels at both lattice ends should have an effective length of $\ell(1+\delta_\text{out})$.~Taking a channel’s hydraulic diameter of $d_2=3.5$mm and $\ell = 4$mm as a case in point, the effective length evaluates to $\ell_e=5.05$mm. Setting up the numerical problem as described above, the overall distribution of the natural frequencies from the numerical solution excellently match those from the analytical model (Figs.~4(\textit{d})~and~5(\textit{a})). This is further emphasized from the mode shapes for the in-gap edge states as shown in Fig.~5(\textit{b}), which shows wave localization at edges with strong wave attenuation away from them, as expected. Figure~5 also evinces the agreement in the natural frequencies/mode shapes of the numerical three-dimensional model with their analytical counterpart. Note that the results in Fig.~5 are obtained based on an identical parameter set to that in Fig.~4(\textit{a}). 

\section{Acoustic honeycomb lattices}

\begin{figure*}[]
     \centering
\includegraphics[]{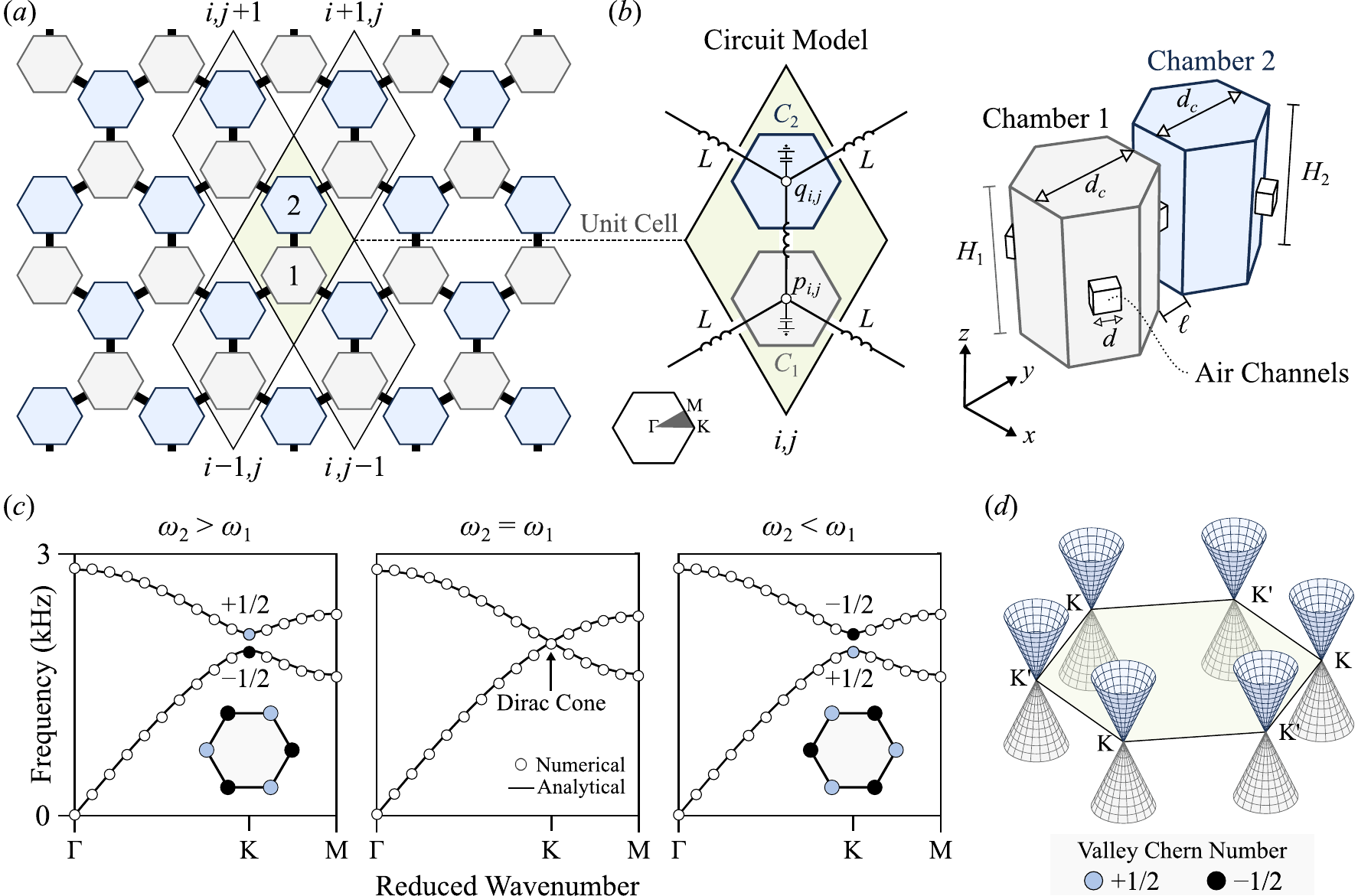}
     \caption{(\textit{a,b}) Schematics of an acoustic honeycomb lattice, the unit-cell definition, and the equivalent circuit model of the unit cell.~Note that the change in the chamber height occurs symmetrically such that the coupling square channels remain at the center of the chambers' height.~(\textit{c}) Dispersion diagrams for $\omega_2>\omega_1$, $\omega_2=\omega_1$, and $\omega_2<\omega_1$, corresponding to $\Delta H=+4\text{mm},0,-4\text{mm}$, respectively, with the average height being $H_0 = 20$mm (See Equation~(\ref{eq:H12})). Other geometrical parameters for simulations remain constant and are as follows: $d = 3$mm, $d_c = 14$mm, and $\ell = 4$mm.~Valley Hall transition occurs exactly at $\Delta H=0$ (when a Dirac cone is generated) and it is dictated by flipping the sign of the valley Chern number (blue circles are $+1/2$ and black circles are $-1/2$).~(\textit{d}) The linearized Hamiltonian at the corners of the Brilloiun zone, exhibiting perfect cones at K (K') points.}
     \label{fig:HoneycombLattice_disp}
\end{figure*}

\subsection{Dispersion surfaces and unit-cell Hamiltonian}
Next, consider an acoustic honeycomb lattice and the equivalent circuit model of its unit cell (Fig.~6(\textit{a},\textit{b})). Chamber pressure of the first and second cavity of the $(i,j)^{\text{th}}$ unit cell are denoted as $p_{i,j}$ and $q_{i,j}$, respectively. The impedance of the coupling channels is $\mathbf{i}\omega L$, with $L$ being their equivalent inductance and is constant throughout (Fig.~6(\textit{b})). The two chambers of the unit cell, on the other hand, are assigned different heights $H_1$ and $H_2$, and are parameterized as follows:
\begin{equation}
    H_{1,2} = H_0 \pm \frac{\Delta H}{2}
    \label{eq:H12}
\end{equation}
where $+$ ($-$) sign correspond to $H_1$ ($H_2$). Here, $H_0$ is the average of the heights and $\Delta H$ is the total difference between them.~On account of the differing heights $H_1$ and $H_2$, distinct equivalent capacitance of $C_1$ and $C_2$ for the different unit-cell chambers is expected. As a result, the governing equations of a unit cell can be shown to be:
\begin{subequations}
\begin{equation}
    \left(\frac{3}{\mathbf{i}\omega L} + \mathbf{i}\omega C_1 \right) p_{i,j}-\frac{1}{\mathbf{i}\omega L}\left(q_{i,j}+ q_{i-1,j} + q_{i,j-1} \right)=0
\end{equation}
\begin{equation}
\left(\frac{3}{\mathbf{i}\omega L} +\mathbf{i}\omega C_2 \right) q_{i,j}-\frac{1}{\mathbf{i}\omega L}\left(p_{i,j}+ p_{i+1,j} + p_{i,j+1} \right)=0
\end{equation}
\label{eq:Honeycomb_EOM}
\end{subequations}
Applying Bloch theorm and introducing,
\begin{subequations}
\begin{equation}
\omega_{1,2}=\sqrt{1/(LC_{1,2})}
\end{equation}
\begin{equation}
    \varepsilon = 1 + \text{e}^{\mathbf{i}\mu_+} + \text{e}^{\mathbf{i}\mu_-}
\end{equation}
\begin{equation}
    \mu_{\pm} = \frac{1}{2}\left(\sqrt{3} \mu_y \pm  \mu_x \right)
\end{equation}
\end{subequations}
the degrees of freedom in Equation (\ref{eq:Honeycomb_EOM}) can be condensed and cast into a similar matrix form to Equation~(\ref{eq:EVP_DL}) with:
\begin{subequations}
\begin{equation}
\mathbf{p}_{i,j}=
    \begin{Bmatrix}
    p_{i,j} & q_{i,j}
    \end{Bmatrix}^{\text{T}}
\end{equation}
\begin{equation}
\mathbf{H} = 
    \begin{bmatrix}
    3\omega_1^2 & -\omega_1^2 \varepsilon^\dagger \\
    -\omega_2^2 \varepsilon & 3\omega_2^2
    \end{bmatrix}
    \label{eq:H_honeycomb1}
\end{equation}
\end{subequations}
The dimensionless wavenumbers $\mu_\pm$ are a function of $\mu_x$ and $\mu_y$, defined as the wavenumbers in the $x$- and $y$- directions, respectively. It is observed that the Hamiltonian in Equation~(\ref{eq:H_honeycomb1}) is not symmetric, thus cannot be expressed in terms of Pauli matrices. To achieve such a symmetric  Hamiltonian and properly interpret topological properties, a new basis $\mathbf{\hat{p}}= \mathbf{Q}\mathbf{p}_{i,j}$ is introduced, parallel to the methodology presented in Ref.~\cite{pal2017edge}, with the definition of the transformation matrix $\mathbf{Q}$ being:
\begin{equation}
    \mathbf{Q} = \mathbf{diag}
    \begin{bmatrix}
    1/\omega_1 & 1/\omega_2
    \end{bmatrix}
    \label{eq:Q}
\end{equation}
As such, we arrive at the following eigenvalue problem,
\begin{equation}
    \mathbf{\hat{H}}\mathbf{\hat{p}} = \omega^2 \mathbf{\hat{p}} 
\end{equation}
where $\mathbf{\hat{H}} = \mathbf{Q}\mathbf{H}\mathbf{Q}^{-1}$ and it is given by:
\begin{equation}
\mathbf{\hat{H}} = 
    \begin{bmatrix}
    3\omega_1^2 & -\omega_1\omega_2 \varepsilon^\dagger \\
    -\omega_1\omega_2 \varepsilon & 3\omega_2^2
    \end{bmatrix}
    \label{eq:H_honeycomb}
\end{equation}
The Hamiltonian in Equation~(\ref{eq:H_honeycomb}) can be now expressed as a summation of Pauli matrices, i.e., Equation~(\ref{eq:Pauli_matrices}), with the following parameters: 
\begin{subequations}
\begin{align}
    h_0 &= -3 (\omega_1^2 + \omega_2^2)/2 \\
    h_1 &= \omega_1 \omega_2 \left(1+\cos(\mu_+)+\cos(\mu_-) \right) \\
    h_2 &= \omega_1 \omega_2 \left(\sin(\mu_+)+\sin(\mu_-) \right) \\
    h_3 &= -3(\omega_1^2 - \omega_2^2)/2
    \label{eq:h3_HC}
\end{align}
\end{subequations}
By defining $\text{E}(\mu_+,\mu_-)=\left(\cos(\mu_+)+\cos(\mu_-)+\cos(\mu_+-\mu_-)-3 \right)$, the associated eigenvalues of $\mathbf{\hat{H}}$ (or equivalently $\mathbf{H}$ in Equation~(\ref{eq:H_honeycomb1})) dictating the dispersion surfaces are expressed in the following compact form:
\begin{equation}
    \omega = \sqrt{\frac{3}{2}\left(\omega_1^2 + \omega_2^2 \right) \pm \sqrt{\frac{9}{4}(\omega_1^2 + \omega_2^2)^2 + 2 \omega_1^2 \omega_2^2 \text{E}(\mu_+,\mu_-)} }
    \label{eq:disp_HC}
\end{equation}

For $H_0 = 20$mm, $d = 3$mm, $d_c = 14$mm, and $\ell = 4$mm, Fig.~6(\textit{c}) shows the dispersion diagrams for $\omega_2>\omega_1$, $\omega_2=\omega_1$, and $\omega_2<\omega_1$, corresponding to $\Delta H=+4\text{mm},0,-4\text{mm}$, respectively (See Equation~(\ref{eq:H12})). On a first glance, the cases of $\Delta H=\pm4\text{mm}$ look identical, a fact that is observed in both the numerical and analytical models which are in excellent agreement. However, valley Hall transition takes place at $\Delta H=0$ (when a Dirac cone is generated) and it is dictated by flipping the sign of the valley Chern number as will be established shortly. A complete bandgap only opens if $\Delta H \neq 0$ and its limits are found from the solutions of the dispersion relation in (\ref{eq:disp_HC}) at K point (or equivalently K' point), i.e., $(\mu_x^\text{K},\mu_y^\text{K})=(4\pi/3,0)$, resulting in $\omega =\sqrt{3} \omega_{1,2}$. 

\begin{figure*}[ht]
     \centering
\includegraphics[]{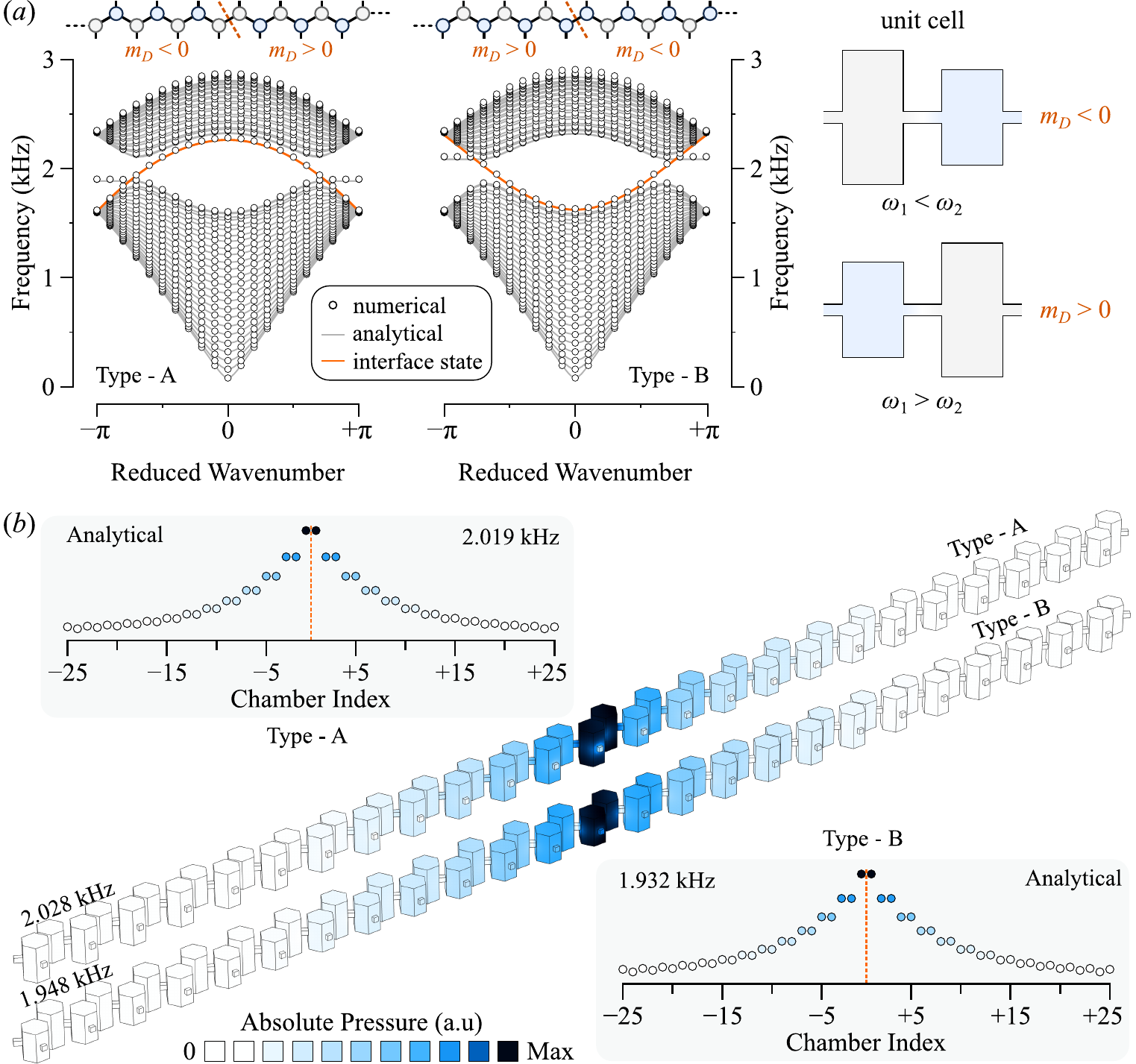}
     \caption{(\textit{a}) Supercell analysis for a honeycomb lattice, formed by merging two lattices with flipped order of chambers (i.e., sign of $m_D$) at an interface midway, with 25 chambers on each side.~The spectrum based on the full-scale numerical model (circles) and analytical one (lines) are shown for two types of interface (orange lines): Type-A (large chambers interface) and Type-B (small chambers interface). (\textit{b}) Corresponding eigenmodes at reduced wavenumber of $0.6\pi$ based on the numerical and analytical models.}
     \label{fig:HoneycombLattice_strip}
\end{figure*}

\subsection{Valley Chern number}

A key parameter in understanding valley Hall transition is $h_3$ from Equation~(\ref{eq:h3_HC}) and its presence results in breaking the inversion symmetry and opens a frequency bandgap.~Breaking such inversion symmetry and maintaining a third order rotational symmetry give rise to QVHE~\cite{pal2017edge,AlBabaa2020Elastically-supportedInsulators}. The topological invariant quantifying the topological transition is the valley Chern number $C_v$, which is calculated based on a linearized version of the Hamiltonian in Equation~(\ref{eq:H_honeycomb}) near a Dirac cone.~The latter is simply achieved by expanding $\mathbf{\hat{H}}$ via a multi-variable Taylor series near the K point, while ignoring the constant diagonal entries (i.e., $h_0 \boldsymbol{\sigma}_0$),
\begin{equation}
\delta\mathbf{\hat{H}} = \mathbf{\hat{H}}(\mu_x^\text{K},\mu_y^\text{K}) + \left(\partial_{\mu_x}\mathbf{\hat{H}}\Big|_\text{K}\right)\delta \mu_x  + \left(\partial_{\mu_y}\mathbf{\hat{H}}\Big|_\text{K}\right) \delta \mu_y
\end{equation}
where $\delta \mu_x$($\delta \mu_y$) are wavenumbers measured from $\mu_x^{\text{K}}$($\mu_y^{\text{K}}$).~As a result, the diagonal elements of $\delta\mathbf{\hat{H}}$ is identical to $h_3 \boldsymbol{\sigma}_3$, which emanates from the first term in the expansion.~The off-diagonal elements, on the other hand, are found via expanding the function $\varepsilon$ near the K-point as follows:
\begin{equation}
    \delta \varepsilon = \varepsilon(\mu_x^\text{K},\mu_y^\text{K}) + \frac{\partial \varepsilon}{\partial{\mu_x}}\Big|_\text{K} \delta \mu_x + \frac{\partial \varepsilon}{\partial{\mu_y}} \Big |_\text{K} \delta \mu_y
    \label{eq:linearized_vareps}
\end{equation}
An analogous procedure can be followed for the complex conjugate $\varepsilon^\dagger$.~Evaluating the expression at the K point and knowing that $\varepsilon(\mu_x^\text{K},\mu_y^\text{K})=0$, Equation (\ref{eq:linearized_vareps}) boils down to:
\begin{equation}
    \delta \varepsilon = -\frac{\sqrt{3}}{2} \left(\tau \delta\mu_x + \mathbf{i} \delta\mu_y \right)
\end{equation}
where $\tau = +1(-1)$ for the K(K') point. Subsequently, the linearized Hamiltonian can be written as~\cite{Tian2020DispersionCrystals}:
\begin{equation}
    \delta \mathbf{\hat{H}} = v_D (\tau \delta \mu_x \boldsymbol{\sigma}_1+\delta \mu_y \boldsymbol{\sigma}_2)+m_D v_D^2 \boldsymbol{\sigma}_3
\end{equation}
where $v_D$ and $m_D$ are the Dirac velocity and effective mass, respectively, and, for our configuration, are given by:\begin{subequations}
\begin{align}
    v_D &= \frac{\sqrt{3}}{2} \omega_1 \omega_2 \\
    m_D &= 2 \frac{(\omega_1^2 -  \omega_2^2)}{(\omega_1 \omega_2 )^2} 
\end{align}
\end{subequations}
Based on the linearized Hamiltonian $\delta \mathbf{\hat{H}}$, the following eigenvalue problem
\begin{equation}
    \delta \mathbf{\hat{H}} \delta \mathbf{\hat{p}} = \delta \omega^2 \delta \mathbf{\hat{p}}
\end{equation}
yields an eigenpair of the form:
\begin{subequations}
\begin{equation}
    \delta \omega = \pm v_D \sqrt{(m_D v_D)^2 + \delta \mu_x^2 + \delta \mu_y^2}
    \label{eq:EValue_linear_H}
\end{equation}
\begin{equation}
     \delta \mathbf{\hat{p}} =
     \begin{Bmatrix}
     m_D v_D^2 \pm |\delta\omega^2| &
     v_D \left(\tau \delta \mu_x + \mathbf{i} \delta \mu_y \right)
     \end{Bmatrix}^{\text{T}}
     \label{eq:EVector_linear_H}
\end{equation}
\end{subequations}
It is important to point out that if $m_D=0$, i.e., $\omega_1^2=\omega_2^2=\omega_0^2$, the eigenvalues in Equation (\ref{eq:EValue_linear_H}) describe perfect cones:
\begin{equation}
    \delta \omega = \pm \frac{1}{2} \omega_0 \sqrt{3 (\delta \mu_x^2 + \delta \mu_y^2)}
\end{equation}
and are depicted in Fig.~6(\textit{d}) for reference.~Such a result has been similarly established for an in-plane Kagome elastic lattice \cite{chen2018topological}. The eigenvectors in Equation (\ref{eq:EVector_linear_H}) are crucial to calculate the Berry curvature, which, after normalizing $\delta \mathbf{\hat{p}}$, can be shown to be~\cite{Tian2020DispersionCrystals}:
\begin{equation}
    \mathcal{F}_\pm = \frac{\mp \tau m_D v_D}{2(m_D^2 v_D^2 + \delta \mu_x^2 + \delta \mu_y^2)^{3/2}}
\end{equation}
Finally, the integration of Berry curvature near a single valley yields a quantized valley Chern number:
\begin{equation}
    C_v^{\pm} = \mp \frac{1}{2} \mathbf{sgn}[m_D]
\end{equation}

\begin{figure*}[]
     \centering
\includegraphics[]{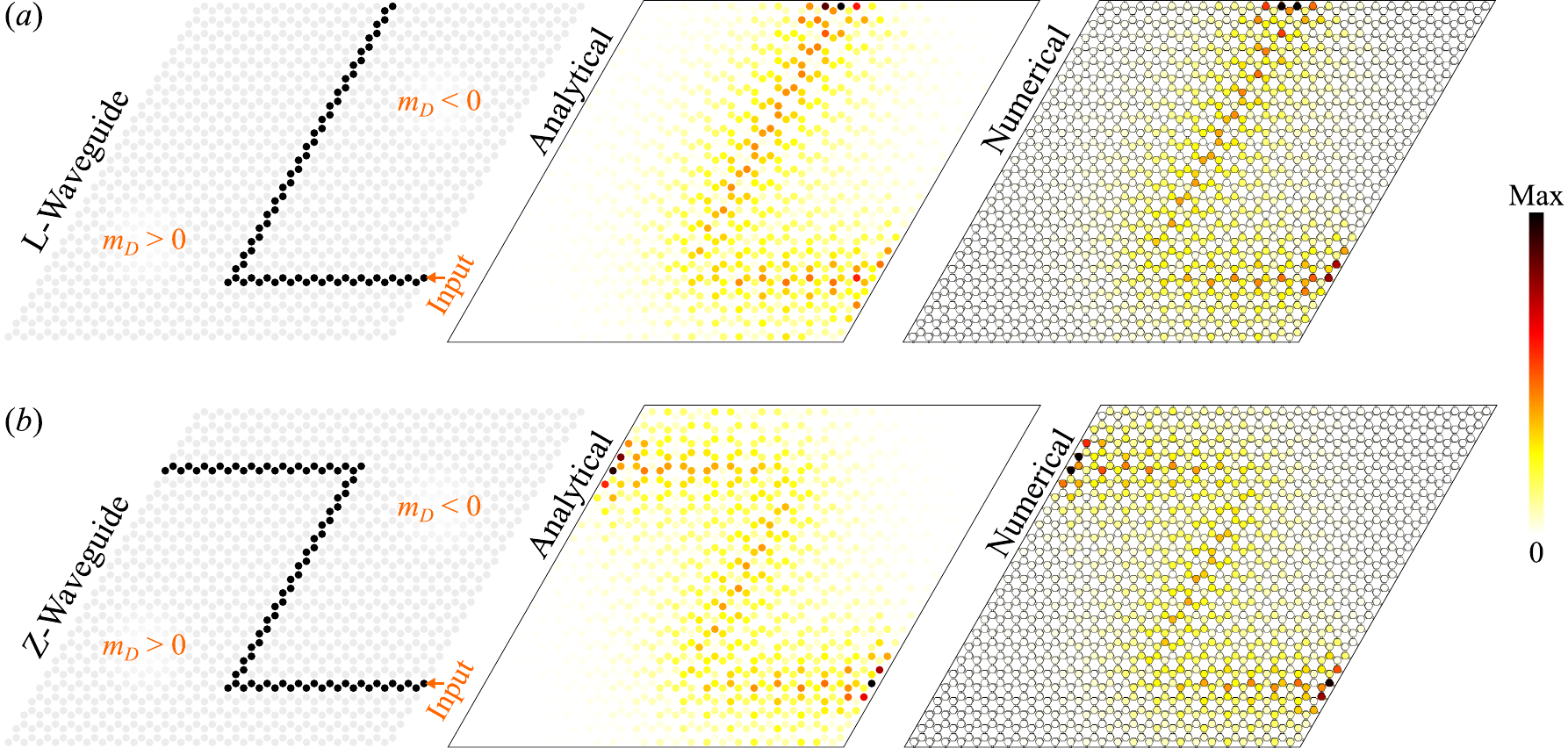}
     \caption{A parallelogram shaped honeycomb structure with embedded (\textit{a}) L- and (\textit{b}) Z-waveguides, which are made of Type-B interface, defined in Fig.~\ref{fig:HoneycombLattice_strip}.~An arbitrary input is imposed on the right end of the waveguides at a frequency of 2kHz (around the center of the frequency bandgap). The numerical and analytical models show qualitatively similar behavior.}
     \label{fig:HoneycombLattice_finite}
\end{figure*}

\subsection{Finite lattice dynamics and waveguide design}
One of the implications of having a topological transition is to design robust interface states and waveguides.~This can be demonstrated by performing standard supercell dispersion analysis on a single strip of the honeycomb lattice with flipped order of the lattice chambers at a midway interface with the second dimension being infinite (Fig.~7(\textit{a})). This configuration results in a lattice with two parts having opposite signs of $C_v$ (or opposite signs of $m_D$), and, consequently, there should exist a single in-gap interface state~\cite{Riva2018TunableLattices}.~Performing such analysis in a chain of 50 chambers with parameters identical to that of Fig.~6(\textit{c}), an interface state (depicted in orange in Fig.~7(\textit{a})) is observed in both the analytical and numerical (three-dimensional) models, with an overall good agreement. A~distinguishing feature between the solo interface state arising from an interface of Type-A and Type-B is the group velocity (i.e., slope of the curve), which flips its sign with changing the interface type.~The nature of such interface states is further confirmed by examining the mode shape; for example, Fig.~7(\textit{b}) shows the absolute pressure of a mode at a reduced wavenumber of $0.6\pi$, which, once again, shows an excellent agreement between the analytical and numerical models. Note here that the left/right boundaries are terminated with an open condition on both lattice’s ends and the length of peripheral channels are adjusted in a similar way to that of the diatomic lattice in section~\ref{sec:DL_edge_states}. 

Based on the supercell analysis, waveguides with L- and Z- shapes in a parallelogram-like lattice are devised using Type-B interfaces (Fig.~8).~The acoustic lattice is built from $n = 25$ rows of supercell lattice strips defined in Fig. 7(\textit{b}), with each strip having an $n$ unit cells, to form a parallelogram-like honeycomb lattice with a total of $n_t=2n^2 = 1250$ chambers. Examining the system's response to an arbitrary excitation of 2kHz at the right end of both waveguides, it is evident that the pressure is localized at the waveguide as seen in Fig.~8(\textit{a},\textit{b}), precisely as predicted by the supercell analysis in Fig.~7(\textit{b}). Both the numerical and analytical models are in a nearly perfect agreement, which demonstrates the power of the analytical model, while having less total number of degrees of freedom, all without any apparent difference in the overall response. 

\section{Conclusions}
This paper has established the use of electroacoustic analogies (i.e., mapping acoustic components to electrical circuit elements) for designing topological acoustic lattices.~The physical meaning of the parameters established from the effective Hamiltonian is related to Helmholtz resonance, allowing for a better estimation of its frequency based on geometrical properties. The benefit of the established platform is its smaller demand of numerical computation, in comparison to a full-scale three-dimensional model.~Two examples of topological acoustic systems have been studied to demonstrate the effectiveness of such approach:~(i) an acoustic diatomic lattice with topological edge states, and~(ii) an acoustic honeycomb lattice with embedded topological waveguides emanating from QVHE. Both cases show excellent agreement between the analytical circuit model and the full-scale three-dimensional one, whether in finite-lattice frequency response or unit-cell based analyses.~It is also emphasized that the predicting accuracy of acoustic-lattice resonances is susceptible to the chosen correction factors of the channel's length at the boundaries of the three-dimensional model. The simplicity and computational efficiency of the established framework can be invaluable for designing various topological acoustic lattices in the future. 

\section*{Acknowledgement}
The authors acknowledge  funding from the U.S.
Air Force Office of Scientific Research (FA9550-18-1-0192).
\newpage
\section*{References}
\begin{multicols}{2}
\footnotesize
\printbibliography[heading=none]
\end{multicols}
\break
\end{document}